\documentstyle{aipproc}

\begin{document}
\def\be{\begin{equation}}
 \def \ee{\end{equation}}   
\def\bea{\begin{eqnarray}}\def\eea{\end{eqnarray}}
  \title{Nuclear Physics on the Light Front--a new old way to do an old new
  problem}

\author{Gerald A. Miller$^*$}
                                
\address{$^*$Department of Physics\\
  University of Washington\\
  Seattle, Washington 98195-1560\thanks{This work is partially supported by
    the  USDOE}}
\maketitle

\begin{abstract}
A brief introduction to light front techniques is presented. This is followed
by a review of recent attempts to perform realistic, relativistic nuclear
physics with those techniques.
\end{abstract}

\section*{Motivation}

This lecture series  is aimed at  describing our recent attempts to
derive the properties of  nuclei using  the light-front formalism.
 Nuclear
properties are very well handled within  existing conventional
nuclear theory, so it is necessary to
explain the motivation.  It seems to me  that understanding experiments
involving high energy nuclear reactions  requires that
light-front dynamics and light cone variables be used. Consider the 
EMC experiment \cite{emc}, which showed that there is a significant
difference between the parton distributions of free nucleons and
nucleons in a nucleus. This difference can interpreted as a shift in 
the momentum distribution of valence quarks towards smaller values of
the Bjorken variable $x$. This  variable is a ratio
of the plus-momentum $k^+=k^0+k^3$ of a quark to that of the target.
If one uses $k^+$ as a momentum
variable, the corresponding canonical spatial variable is $x^-=x^0-x^3$
and the time variable is $x^+=x^0 +x^3$. To do calculations
in this framework is
to use light front dynamics.

Light front dynamics applies to nucleons
within the nucleus as well as to partons of the nucleons, and 
this is a useful approach whenever  the momentum of
initial or final state nucleons is large compared to their
mass \cite{fs}. For example,
this technique  can be used for $(e,e'p)$ and $(p,2p)$
reactions at sufficiently high energies. The use of
 light-front variables for nucleons in a nucleus is not sufficient.
It is also necessary to include all the relevant features of
 conventional nuclear dynamics. Combining these two aspects  provides
the technical challenge which we have been addressing.

I'd like to begin by describing how  using the
light-front approach leads to important simplifications.
Consider high energy electron scattering from nucleons in nuclei.
Let the four-momentum $q$ of the exchanged virtual photon be
given by $\left(\nu,0,0,-\sqrt{Q^2+\nu^2}\right)$, with $Q^2=-q^2$, and
$Q^2$ and $\nu^2$ are both very large but $Q^2/\nu$ is finite (the
Bjorken limit). Use the light-cone
variables $q^\pm=q^0\pm q^3$ in which $q^+\approx Q^2/2\nu=Mx$,
$q^-\approx 2\nu -Q^2/2\nu$, so that $q^-\gg q^+$. Here $M$ is the mass
of a nucleon.  We  neglect $q^-$ in
comparison to $q^+$;  corrections to this can be handled in
a systematic fashion. Then the  scattering cross
section for $e+A\to e' +(A-1)_f +p$, where $f$ represents the final
nuclear eigenstate of $P^-$, and $p$ the four-momentum of the final
proton, takes the form
\begin{equation}
d\sigma\sim \sum_f \int
{d^3p_f\over E_f}\int d^4p\, \delta (p^2-M^2)\delta^{(4)}(q+p_i-p_f-p)|
\langle p,f\mid J(q)\mid i\rangle\mid^2,
\end{equation}
with  the operator $J(q)$  as a schematic representation of the
electromagnetic current. Performing the four-dimensional integral over
$p$ leads to the expression
\begin{equation}
d\sigma\sim \sum_f \int {d^2p_fdp_f^+\over p^+_f}
\delta\left((p_i-p_f+q)^2-m^2\right)\mid \langle p,f\mid J(q)\mid
i\rangle\mid^2 \label{int}.
\end{equation}
The argument of the delta function $(p_i-p_f+q)^2-M^2 \approx
-Q^2+2q^-(p_i-p_f)^+$. Thus  $p_f^-$ does not appear in the
argument of the delta function, or anywhere else, so that we can
replace the sum over intermediate states by unity. In the usual
equal-time representation,  the argument of the delta function
is  $-Q^2+2\nu(E_i-E_f)$. The energy of the final state appears, and
one can not do the sum over states.
It is useful to define $\bbox{p_B}\equiv
\bbox{p_i}-\bbox{p_f}$,  because
$ 
p_B^+=Q^2/2\nu\equiv M x  
$
(as demanded by the delta function). Then one can re-express
Eq.~(\ref{int}) as
\begin{equation}
d\sigma\sim
 d^2{p_B}_\perp n(M x,{p_B}_\perp),
\label{ds2}
\end{equation}
where $ n(M x,{p_B}_\perp)$ is the probability for a nucleon in
the ground state to have a momentum $(M x,{p_B}_\perp)$. Integration in
Eq.~(\ref{ds2}) leads to
\begin{equation}
\sigma \sim\int d^2p_\perp\, n(M x,{p}_\perp)\equiv f(Mx),
\end{equation}
with $f(Mx)$ as the probability for a nucleon in the ground state to
have a plus momentum of $Mx$. 
The use of light-front dynamics to compute nuclear wave functions
should allow us to compute $f(Mx)$ from first principles. 
We also claim that using light-front dynamics incorporates the
experimentally relevant kinematics from the beginning, and therefore is
the most efficient way to compute the cross sections for nuclear deep
inelastic scattering and nuclear quasi-elastic scattering. 

Since much of this
work is
motivated by the desire to understand nuclear deep inelastic scattering
and related experiments,
it is worthwhile to review some of the features of the EMC
effect \cite{emc,emcrevs}. One key experimental result is the
suppression of the structure function for $x\sim 0.5$. This means that
the valence quarks of bound nucleons carry less plus-momentum than
those of free nucleons. This may be understood by
postulating that mesons carry a larger fraction of the plus-momentum in
the nucleus than in free space. While such a model explains the shift
in the valence distribution, one obtains at the same time a meson (i.e.
anti-quark) distribution in the nucleus, which is strongly enhanced
compared to free nucleons and which should be observable in Drell-Yan
experiments \cite{dyth}. However, no such enhancement has been observed
experimentally \cite{dyexp}, and the implications are analyzed in
Ref.~\cite{missing}.

The use of light-front dynamics should allow us to compute the necessary
nuclear meson distribution functions using variables which are
experimentally relevant. The need for a computation of such functions
in a manner consistent with generally known properties of nuclei led
us to begin this program.
There are  other motivations for using the light front formalism that have
been emphasized in many reviews\cite{lcrevs}.
One key feature is that the vacuum of the
theory is trivial because it can not create pairs.  Another is
that the theory is a Hamiltonian theory and the many-body techniques of 
equal time theory can be used here too.
I also like to say: Ask not what the light front can do for nuclear physics;
instead
ask
what nuclear physics can do for the light front. This is to provide a set of
non-trivial four dimensional examples with real physics content.
Finally I quote the review by Geesaman
et al.
``In light front dynamics LFD,
the
particles are on mass-shell, and there are no off-shell
ambiguities. However, ... we have little or
no experience in
calculating the wave function of a  realistic nucleus in LFD''.
The aim
here is to provide such wave functions. 
\subsection*{Outline}
We shall 
begin with a simple description of what is light front dynamics. Then
the formal procedures of light front quantization of a hadronic Lagrangian
${\cal L}$
will be discussed. The first application is a study of 
infinite nuclear matter within the mean field approximation\cite{jerry}.
The distribution
functions $f(y)$ for nucleons and mesons will be computed.
The above topics comprise the first lecture. The next lecture is devoted
to a study of
finite nuclei\cite{bbm99}
using the  mean field approximation. Here one must confront
a difficulty. The use of $x^-=t-z$ as a spatial variable violates manifest
 rotational
invariance because  $x^-$ and $x_\perp$ are different variables.
We show that rotational invariance re-emerges after one does the appropriate
dynamical calculation.
It is necessary to go beyond the mean field approximation, and  the third
lecture
deals with that\cite{rmgm98}. Nucleon-nucleon scattering is   studied
first and used in the many-body calculation.  The influence of 
nucleon-nucleon correlations on the properties of nuclear matter is studied by
making the necessary light front calculations. Applications are to compute the
nuclear pionic content and to nuclear deep inelastic scattering and 
Drell-Yan processes.
The  goal is to provide a series of 
examples showing  that the light front  can be
used for high energy realistic and relativistic nuclear physics.

\section*{ What is light front dynamics? }

This is a relativistic treatment
of dynamics in which the fields are quantized at a fixed ``time''$\tau =t+z
=x^0+x^3\equiv x^+$. This means that the orthogonal spatial variable must be 
$x^-\equiv t-z$ so that the canonical momentum is $ p^0+p^3\equiv p^+$. The
remainder of the spatial variables are given by:
$ \vec{x}_\perp,\vec{p}_\perp$.

The consequence of using $\tau$ as a 
 ``time'' variable is that the canonical energy is $p^-=p^0-p^3$.
 In general our notation is given by
 \begin{equation}
A^\pm\equiv A^0\pm A^3,
\end{equation}
with
\begin{equation}
A\cdot B =A^\mu B_\mu={1\over2}\left(A^+B^- +A^-B^+\right)
-\vec{A}_\perp\cdot\vec{B}_\perp
.\end{equation}

The key reason for using such unusual coordinates is phenomenological.
 For a particle with  $\vec v\approx c\hat{e_3}$, the quantity   $p^+$
is BIG. Thus experiments tend to measure quantities associated with
$p^+$.

Another important feature is the relativistic dispersion relation
$p^\mu p_\mu =m^2$, which
in light front dynamics takes the form:
 \begin{equation}
 p^-={p_\perp^2+m^2\over p^+}
 .\end{equation}
Thus one has a form of relativistic kinematics that avoids using a
square root.

The main formal consequence of using light front dynamics is that the minus
component of the total momentum,
$P^-$, is used as a Hamiltonian operator, and the plus component 
 $P^+$ is used as a momentum  operator.  The procedures to obtain these
 operators are discussed in the next section.

\section*{Light Front Quantization}

My intent here is to discuss the basic aspects in as informal way as possible.
For more details see the reviews and the references. I'll start by considering
one free field at a time. These will be the scalar meson $\phi$, the Dirac
fermion $\psi$ and  the massive vector meson $V^\mu$. 
\subsection*{Free Scalar field}
Consider the Lagrangian
\begin{eqnarray}
{\cal L}_\phi 
= {1\over 2} (\partial^+\phi \partial^-\phi -\bbox{\nabla}_\perp\phi
\cdot\bbox{\nabla}_\perp\phi-m_s^2\phi^2).\label{lagphi}
\end{eqnarray}
The notation is such that
$
  \partial^\pm=\partial^0\pm\partial^3=2 {\partial\over \partial x^\mp}$.
The Euler-Lagrange equation
leads to the wave equation
\begin{eqnarray}
  i\partial^-\phi={-\nabla^2_\perp+m_s^2\over i\partial^+}\;\phi.
\end{eqnarray}
The most general solution is a superposition of plane waves:
\begin{equation}
\phi(x)=
\int{ d^2k_\perp dk^+ \theta(k^+)\over (2\pi)^{3/2}\sqrt{2k^+}}\left[
a(\bbox{k})e^{-ik\cdot x}
+a^\dagger(\bbox{k})e^{ik\cdot x}\right],
\label{expp}\end {equation} 
where 
$k\cdot x={1\over2}(k^-x^++k^+x^-)-\bbox{k_\perp\cdot x}_\perp$ with
 $k^-={k_\perp^2+m_s^2\over k^+}$, and $\bbox{k}\equiv(k^+,\bbox{k}_\perp)$.
The $\theta$ function restricts $k^+$ to positive values.
Note that
\begin{equation}
  i\partial^+e^{-ik\cdot x}=k^+e^{-ik\cdot x}
      .\end{equation}
    The value of $x^+$ that appears in Eq.~(\ref{expp}) can be set to zero, but
    only after taking necessary derivatives.

    Deriving the equal $x^+$ commutation relations for the fields
    is a somewhat obscure
    procedure \cite{yan12}, but the result can be stated in terms of  familiar
commutation relations:
\begin{equation}
[a(\bbox{k}),a^\dagger(\bbox{k}')]=
\delta(\bbox{k}_\perp-\bbox{k}'_\perp)
\delta(k^+-k'^+)\label{comm}
\end {equation}
with $[a(\bbox{k}),a(\bbox{k}')]=0$.

The next step is compute the Hamiltonian $P^-$ for this system.
The conserved  energy-momentum tensor is given in terms of the Lagrangian:
\begin{equation}
T^{\mu\nu}_\phi=-g^{\mu\nu}{\cal L_\phi} +{\partial{\cal L}_\phi
  \over\partial
  (\partial_\mu\phi)}\partial^\nu\phi.\label{tmunup}
.\end {equation}
This brings us to the question of what is $g^{\mu\nu}$?
This is straightforward, although the results (viewed for the first time)
can be surprising:
\begin{eqnarray}
  g^{+\nu}=g^{0\nu}+g^{3\nu}
  \end{eqnarray}
  Thus
 \begin{eqnarray} g^{++}&=&g^{00}+g^{03}+g^{30}+g^{33}=1+0+0-1=0\nonumber\\
 g^{ij}&=&-\delta_{i,j} (i=1,2,j=1,2);\quad
g^{+-}=g^{-+}=2.
\end{eqnarray}

Then one finds that
\begin{equation}
T^{+-}_\phi={1\over 2}\bbox{\nabla}_\perp \phi \cdot \bbox{\nabla}_\perp\phi+
{1\over 2}m^2_s \phi^2.
\end {equation}
The term $T^{+-}$  is the density for the operator $P^-$:
\begin{equation}
P^-={1\over 2}\int d^2x_\perp dx^- T^{+-}.
\end{equation}
The use of the field expansion (\ref{expp}), along with normal ordering
followed by integration leads to the result:
\begin{equation}
P^-_\phi=\int d^2k_\perp dk^+
\theta(k^+)a^\dagger(\bbox{k})a(\bbox{k}){k_\perp^2+m_s^2\over k^+}.
\end{equation}
One defines a vacuum state $\mid0\rangle$ such that
$
  a(\bbox{p})\mid0\rangle=0.
$
Then the creation operators acting on the vacuum give the usual single particle
states:
\begin{eqnarray}
  P^-_\phi a^\dagger(\bbox{p})\mid0\rangle={p_\perp^2+m_s^2\over p^+}
  a^\dagger(\bbox{p})\mid0\rangle.\end{eqnarray}
The momentum operator $P^+$ is constructed by integrating $T^{++}$:
\begin{equation}
P^+_\phi=\int d^2k_\perp dk^+
\theta(k^+)a^\dagger(\bbox{k})a(\bbox{k}) k^+.
\end {equation}

\subsubsection*{Interactions and Light Front Simplification}
Suppose we take the Lagrangian
\begin{eqnarray}
{\cal L}={1\over 2} (\partial_\mu \phi \partial^\mu
\phi-m_s^2\phi^2)+\lambda \phi^4.
\end{eqnarray}
The operator $\phi$ creates or destroys a particles of plus-momenta
$k^+>0$. Thus a possible term in which $\lambda \phi^4$ term converts the
vacuum $\mid0\rangle$ into a four particle state vanishes by virtue of the
conservation of plus-momentum. The vacuum of $p^+=0$ can not be connected
to four particles, each having a positive $k^+$.
 This vanishing simplifies  Hamiltonian  ($x^+$-ordered perturbation)
 calculations.
\subsection*{Free Dirac Field}
Consider the Lagrangian
\begin{eqnarray}
{\cal L}_\psi=
\bar{\psi}(\gamma^\mu
{i\over 2}\stackrel{\leftrightarrow}{\partial}_\mu
-M)\psi, \label{lagpsi}
\end{eqnarray}
and its equation of motion:
\be
\left(i \gamma^\mu\partial_\mu-M\right)\psi=0
\label{dirac0}  .\ee
A
fermion has spin 1/2, so there can only be two independent degrees of
freedom. The standard Dirac spinor has four components, so two of these must
represent dependent degrees of freedom. In the light front formalism one
separates the independent and dependent degrees of freedom by using projection
operators:
$
\Lambda_\pm\equiv {1\over 2} \gamma^0\gamma^\pm
.$ Then the independent field is $\psi_+=\Lambda_+\psi$ and the
dependent one is  $\psi_-=\Lambda_-\psi$

The Dirac equation (\ref{dirac0}) is re-written as
\be
\left({i\over 2}\gamma^+\partial^- +{i\over2}\gamma^-\partial^+
  +i\bbox{\gamma}_\perp\cdot\bbox{\nabla}_\perp-M\right)\psi=0
\label{dirac1}.\ee
Equations for $\psi_\pm$ can be obtained by multiplying
Eq.~(\ref{dirac1}) on the left by $\Lambda_\pm$: 
\bea
i\partial^- \psi_+=(\bbox{\alpha}_\perp\cdot{ \bbox{\nabla}_\perp\over i}
+\beta M)\psi_-\nonumber\\
i\partial^+ \psi_-=(\bbox{\alpha}_\perp\cdot{ \bbox{\nabla}_\perp\over i}
+\beta M)\psi_+,
\eea
so that the equation of motion of $\psi_+$ becomes
\be
i\partial^- \psi_+=(\bbox{\alpha}_\perp\cdot{ \bbox{\nabla}_\perp\over i}+M)
{1\over i\partial^+}(\bbox{\alpha}_\perp\cdot{ \bbox{\nabla}_\perp\over i}+M)\psi_+
.\ee

One can make the field expansion and determine the momenta in a manner similar
to the previous section.
The key results are
\bea
T^{+-}_\psi
&=&\psi^\dagger _+\left(\alpha_\perp\cdot{\bbox{\nabla}_\perp\over i}
  +\beta M\right){1\over
  i\partial^+} \left(\alpha_\perp\cdot{\bbox{\nabla}_\perp\over i}
  +\beta M\right)\psi_+,
\\
P^-_\psi
&=&\sum_\lambda\int d^2p_\perp dp^+\theta(p^+){p_\perp^2+M^2\over  p^+}
\left[b^\dagger(\bbox{p},\lambda)b(\bbox{p},\lambda)+
  d^\dagger(\bbox{p},\lambda)
  d(\bbox{p},\lambda)\right],
\eea
where  $b(\bbox{p},\lambda),d(\bbox{p},\lambda)$ are nucleon and
anti-nucleon destruction operators.
\subsection*{Free Vector Meson}
The formalism for massive vector mesons was worked out by Soper\cite{des71}
  and later
by Yan\cite{yan34}
using a  different formulation. I generally follow Yan's
approach. The  formalism is  lengthy and detailed in the references, so
 I only state the
minimum. There are  three independent degrees of freedom, 
even though the Lagrangian depends on $V^\mu$ and
$V^{\mu\nu}=
\partial ^\mu V^\nu-\partial^\nu V^\mu$. 
These are chosen to be $V^+$ and $V^{+i}$. 
The other terms $ V^-,V^i,V^{-i}$ and $V^{ij}$ can be 
written in terms of $V^+$ and $V^{+i}$. 
\subsection*{We need a Lagrangian, no matter how bad}

It seems to me that one can not do complete dynamical calculations
using the light front formalism without specifying some Lagrangian.
One starts\cite{jerry} with $\cal L$
and derives field equations. These are used to express
the dependent degrees of freedom in terms of independent ones. One also uses
 $\cal L$ to derive $T^{\mu\nu}$ (as a function of independent degrees of
 freedom) which is used to obtain the total momentum
 operators $P^\pm$. It is $P^-$ that acts as a Hamiltonian operator in the
 light front $x^+$-ordered perturbation theory.

 We start with a Lagrangian containing scalar and vector mesons and nucleons
 $\psi'$.
 This is the minimal Lagrangian for obtaining a caricature of nuclear physics
 because the exchange of scalar mesons provides a medium range attraction
 which can bind the nucleons and the exchange of vector mesons provides the
 short-range repulsion which prevents a collapse.
 Thus we take 
\begin{eqnarray}
{\cal L} &=&{1\over 2} (\partial_\mu \phi \partial^\mu \phi-m_s^2\phi^2) 
-{1\over  4} V^{\mu\nu}V_{\mu\nu} +{m_v^2\over 2}V^\mu V_\mu \nonumber\\
&+&\bar{\psi}^\prime\left(\gamma^\mu
({i\over 2}\stackrel{\leftrightarrow}{\partial}_\mu
-g_v\;V_\mu) -
M -g_s\phi\right)\psi', \label{lag} 
\end{eqnarray}
with the effects of other mesons included elsewhere and below.
The equations of motion are
\begin{eqnarray}
\partial_\mu V^{\mu\nu}+m_v^2 V^\nu&=&g_v\bar \psi'\gamma^\nu\psi'
\label{vmeson}\\
\partial_\mu\partial^\mu \phi+m_s^2\phi&=&-g_s\bar\psi'\psi',
\label{smeson}\\
(i\partial^--g_vV^-)\psi'_+&=&(\bbox{\alpha}_\perp\cdot
(\bbox{p}_\perp-g_v\bbox{V}_\perp)+\beta (M +g_s\phi))\psi'_-\label{nfg0}\\
(i\partial^+-g_vV^+)\psi'_-&=&(\bbox{\alpha}_\perp\cdot
(\bbox{p}_\perp-g_v\bbox{V}_\perp)+\beta (M +g_s\phi))\psi'_+. 
\label{nfg}
\end{eqnarray}
The presence of the interaction term $V^+$ on the left-hand  side of the
second equation presents a problem because one can not easily solve for
$\psi_-$ in terms of $\psi_+$.
This difficulty is handled by using the Soper-Yan
transformation:
\be\psi'=e^{-ig_v\Lambda(x)}\psi ,\qquad
\partial^+ \Lambda=V^+. \ee
 Using this in Eqs.~(\ref{nfg0})-(\ref{nfg}) leads to the more usable form
\begin{eqnarray}
(i\partial^--g_v \bar V^-)\psi_+=(\bbox{\alpha}_\perp\cdot 
(\bbox{p}_\perp-g_v\bbox{\bar V}_\perp)+\beta(M+g_s\phi))\psi_-\nonumber\\
i\partial^+\psi_-=(\bbox{\alpha}_\perp\cdot 
(\bbox{p}_\perp-g_v\bbox{\bar V}_\perp)+\beta(M+g_s\phi))\psi_+. 
\label{yan}
\end{eqnarray}
The cost of the transformation is that one gets new terms resulting from
taking derivatives of $\Lambda(x)$. One uses  $\bar V^\mu$ with 
$
 \bar V^\mu=V^\mu-{1\over\partial^+}\partial^\mu V^+,
$
and 
  $\bar V^\mu$
enters in the nucleon field equations, but  
 $V^\mu$  enters in the meson field equations.

\section*{   nuclear matter Mean Field Theory}

The philosophy\cite{bsjdw} is that the nucleonic densities which
are  mesonic sources
are large enough to generate a  large number of
 mesons to enable a classical treatment (replacing an operator by an
 expectation value).  In infinite nuclear matter, the volume is taken
 as infinity so that all positions are equivalent. 
Thus we make the replacement:
\be g_s\bar\psi(x)\psi(x)\to
 g_s\langle\bar\psi(0)\psi(0)\rangle,
\quad\phi={-g_s \over m_s^2}\langle\bar \psi(0)\psi(0)\rangle
,\label{sphi}\ee
in which the expectation value is in the ground state, and second part of the
equation is obtained from the field equation (\ref{smeson}) with a constant
source. Similarly
$g_v\psi(x)\gamma^\mu\psi(x)\to
 g_v\langle\bar{\psi(0)}
\gamma^\mu\psi(0)\rangle 
\delta_{\mu,0},$
in which 
the notion that there is no special direction in space is used.
(The nucleus is taken to be at  rest.)
Again the source is constant,  so that the
solution of the field equation (\ref{vmeson}) is
\be
\bar{V}^-=V^-=V^0={g_v\over m_v^2}\langle\psi^\dagger(0)\psi(0)\rangle;\quad
 \bar{V}^{+,i}=0.
\label{sv}\ee

Since the potentials entering the light-front Dirac equation (\ref{yan}) are
constant, the 
nucleon modes are plane waves $\psi \sim e^{ik\cdot x}$,
and the many-body system is a  kind of
Fermi gas. The solutions of Eq.~(\ref{yan}) are
\be i\partial^- \psi_+=g_v\bar{V}^-\psi_++{k_\perp^2+(M+g_s\phi)^2\over
k^+}\psi_+.\label{sd}\ee
Solving
the equations (\ref{sphi}),(\ref{sv}) and (\ref{sd}) yields
a  self-consistent
solution. 

\subsection*{ Nuclear Momentum Content}
The expectation value of $T^{+\mu}$ is used to obtain the total momentum:
\be P^\mu=
{1\over 2}\int d^2x_\perp dx^-
\langle T^{+\mu} \rangle. \ee The expectation value is constant so that
the volume $\Omega={1\over 2}\int d^2x_\perp dx^- $ will enter as a factor.
A straightforward evaluation leads to the results
\bea
{P^-\over\Omega}&=&m_s^2\phi^2+{4\over
(2\pi)^3}\int_F d^2k_\perp dk^+\;{k_\perp^2+(M+g_s\phi)^2\over k^+}\\
{P^+\over\Omega}&=&m_v^2(V^-)^2+{4\over
(2\pi)^3}\int_F d^2k_\perp dk^+\;k^+.\label{pplus}\eea

To proceed further one needs to define the Fermi surface $F$. The use of a
transformation
$ k^+\equiv \sqrt{(M+g_s\phi)^2+\vec{k}^2} +k^3\equiv E(k)+k^3$
to define a new variable
$ k^3$ enables one to simplify the integrals. One replaces the integral over
$k^+$ by one over $k^3$ (including the Jacobian factor ${\partial k^+\over
  \partial k^3}={k^+\over E})$ 
leads to:
\be\int_F d^2k_\perp dk^+\cdots \equiv \int d^3k\theta(k_F-|\vec k|)\cdots.\ee
The nuclear energy $E$ is the average of $P^+$ and $P^-$:
$
E\equiv{1\over 2}\left(P^-+P^+\right)$ and one gets
the very same expression as in the original Walecka model. This provides a
useful check on the algebra.

There is a potential problem:
for nuclear matter in its rest frame we need to have $P^+=P^-=M_A$. If one
looks at the expressions for $P^\pm$ this result does not seem likely. 
However, the value of the fermi momentum has not yet been determined. There is
one more condition to be satisfied:
\be\left({\partial (E/A)\over\partial k_F}\right)_\Omega=0.\ee
Satisfying this equation determines $k_f$ and for the value so obtained the
values of $P^+$ and $P^-$ turn are the same.

Thus we see that
our light front procedure reproduces standard results for energy and  density.
We use the parameters of Chin and Walecka\cite{cw} $g_v^2M^2/m_v^2=195.9$ and
$g_s^2M^2/m_s^2=267.1$ to obtain first numerical results. Then
$k_F=1.42  \quad$fm$^{-1}$, the binding energy per nucleon is 15.75 MeV and
$M+g_s\phi=0.56
M$. The last number
corresponds to a huge attraction that is nearly cancelled by the
huge repulsion. Then one may use Eq.~(\ref{pplus}) to obtain the separate
contributions of the vector mesons and nucleons, with spectacular  results.
 Nucleons carry only 65\% of the plus-momentum. Thus is much less than
 the  90\% needed to explain the EMC effect for infinite nuclear
 matter\cite{sdm}.
 Furthermore, 
vector mesons carry 35\% of the plus-momentum, which is an amazingly large
number. 

The distribution of this 
vector meson plus-momentum is an interesting quantity. The
mean fields $\phi, V^\mu$  are constants in space and time. Thus 
 $V^-$ has support only for $k^+=0$. The physical interpretation of this
 is that 
$\infty$ number of mesons carry a vanishingly small $\epsilon$ of the
plus-momentum, but the product is  35\%. One can also show\cite{bm98} that
\be k^+f_v(k^+)=0.35 M \delta(k^+).\ee There is an important phenomenological
consequence the value 
$k^+=0$ corresponds to $ x_{Bj}=0$
which can not be  reached in experiments. This means one can't use the
momentum sum rule as a phenomenological tool to analyze deep inelastic
scattering data to determine the different
contributions to the plus-momentum.

Of course this result is caused by solving a simple model for a simple
system with a simple mean field approximation. It is necessary to ask if any
of the qualitative features of the present results will persist in more
detailed treatments.
\section*{Mean Field Theory for finite-sized nuclei}
It is important to make calculations for finite nuclei because all laboratory
experiments are
done for such targets or projectiles. The most basic feature of
all of nuclear physics is that the shell model is able to explain the magic
numbers. Rotational invariance causes the $2j+1$
degeneracy of the single particle orbitals, and full occupation leads to
increased  binding. But light front dynamics does not
make rotational invariance
manifest because the different components of the spatial variable are
treated differently:  $x^-,\bbox{x}_\perp$. However, the final results must
respect rotational invariance. Therefore, the challenge of making  successful
calculations of the properties of finite nuclei is important to us.

Let's 
 discuss, in a general way, how it is that we will be 
able to find spectra which have the correct number of degenerate
states. Suppose we try to determine eigenstates of a LF
Hamiltonian by means of a variational calculation. Simply minimizing
the LF energy  leads to nonsensical results because $P^-=M_A^2/P^+$.
One can easily reach
zero energy by letting $P^+$ be infinite.
This is not a problem if one is able to use a Fock space basis in which
the total plus and $\perp$ momentum of each component are fixed. But in
calculations involving many particles, the Fock state approach cannot
be used in practical calculations. One needs to find a sensible
variational procedure. One such is to 
 perform a constrained variation, in which
the total LF momentum is fixed by including a Lagrange multiplier term
proportional to the total momentum in the LF Hamiltonian.
We minimize the expectation value of $P^+$ subject to the condition that the
expectation values of $P^-$ and $P^+$ are equal. This is
the same as minimizing
the expectation value of the average of $P^-$ and $P^+$.

 The need to
include the plus-momentum along with the minus momentum
can  be seen in a simple example.
Consider a nucleus of $A$ nucleons of momentum $P_A^+=M_A$,
${\bbox{P}_A}_\perp=0$, which consists of a nucleon of momentum
$(p^+,\bbox{p}_\perp)$, and a residual $(A-1)$ nucleon system which
must have momentum $(P^+_A-p^+,-\bbox{p}_\perp)$. The kinetic energy
$K$ is given by the expression
\begin{equation}
K={p_\perp^2+M^2\over p^+}+{p_\perp^2+M_{A-1}^2\over P^+_A- p^+}.
\end{equation}
In the second expression, one is tempted to neglect the term $p^+$ in
comparison with $ P^+_A\approx M_A$. This would be a mistake. Instead
make the expansion
\begin{eqnarray}
K&\approx&{p_\perp^2+M^2\over p^+}+{M_{A-1}^2\over P^+_A}\left(1+ {p^+\over
 P_A^+}\right)\nonumber\\ 
&\approx&{p_\perp^2+M^2\over p^+}+p^+ +M_{A-1}, 
\end{eqnarray}
because for large $A$, $M^2_{A-1}/P_A^2\approx 1$. For free particles,
of ordinary three momentum $\bbox{p}$ one has $E^2(p)=\bbox{p}^2+m^2$
and $p^+=E(p)+p^3$, so that
\begin{equation}
K\approx {\left(E^2(p)-(p^3)^2\right)\over E(p)+p^3}
+E(p)+p^3+M_{A-1}=2E(p)+M_{A-1}.
\end{equation}
We see that $K$ depends only on the magnitude of a three-momentum and
rotational invariance is restored. The physical mechanism of this
restoration is the inclusion of the recoil kinetic energy of the
residual nucleus.

\subsection*{Results}
The formalism is described in  recent papers\cite{bbm99}, so I simply
summarize the results.
If our  solutions  are to have
any relevance,  they should respect rotational invariance. The
success in achieving this is examined in Tables I and II of \cite{bbm99}
  which give
our results for the spectra of $^{16}$O and $^{40}$Ca, respectively.
Scalar and vector meson parameters are taken from Horowitz and
Serot\cite{hs}, and we have assumed isospin symmetry. We see that the
$J_z=\pm1/2$ spectrum contains the eigenvalues of all states, since all
states must have a $J_z=\pm1/2 $ component. Furthermore, the essential
feature that the expected degeneracies among states with different
values of $J_z$ are reproduced numerically. 

The obtained eigenvalues of the nucleon mode equation   are essentially
the same as the single particle energies of the ET
formalism, to within the expected numerical accuracy of our program.
This equality is not mandated by spherical symmetry alone because the
solutions in the equal-time framework have non-vanishing components
with negative values of $p^+$. 
Table III of \cite{bbm99}
gives the contributions to the total $P^+$ momentum from the
nucleons, scalar mesons, and vector mesons for $^{16}$O, $^{40}$Ca, and
$^{80}$Zr, as well as the nuclear matter limit.
The vector mesons carry approximately 30\% of the nuclear
plus-momentum. The technical reason for the difference with the scalar
mesons (which have negligible effect) is that the evaluation of
$a^\dagger(\bbox{k},\omega)a(\bbox{k},\omega)$ counts vector mesons
``in the air''and the resulting expression contains polarization
vectors that give a factor of ${1\over k^+}$  which
enhances the distribution of vector mesons of low $k^+$. The results
for the vector meson distribution are shown in Fig.~2 of \cite{bbm99}. 
As the size of the nucleus increases the enhancement of the
distribution at lower values of $k^+$ becomes more evident. 

\subsection*{Lepton-nucleus deep inelastic scattering}

It is worthwhile to see how the present results are related to 
lepton-nucleus deep inelastic scattering experiments. We find that the
nucleons carry only about 70\% of the plus-momentum. The use of our
$f_N$ in standard convolution formulae lead to a reduction in the
nuclear structure function that is far too large ($\sim$95\% is
needed \cite{emcrevs}) to account for the reduction
observed \cite{emcrevs} in the vicinity of $x\sim 0.5$. The reason for
this is that the quantity $M +g_s\phi$ acts as a nucleon effective mass
of about 670 MeV, which is very small. A similar difficulty occurs in
the $(e,e')$ reaction \cite{frank} when the mean field theory is used
for the initial and final states. The use of a small effective mass and
a large vector potential enables a simple reproduction of the nuclear
spin orbit force \cite{bsjdw,hs}.
Furthermore, the use of other Lagrangians\cite{zim,qmc} will lead to improved
results. We also expect that including  effects beyond the mean field
would  lead to a significant effective tensor coupling of the isoscalar
vector meson \cite{ls}, and to an increased value of the effective mass.
Such effects are incorporated in Brueckner theory, and a light-front
version \cite{rmgm98} could be applied to finite nuclei with better
success in reproducing the data.
This is discussed in the next sections.

\section*{Correlated Infinite Nuclear matter }
The first step is to derive a light front version of the
nucleon-nucleon interaction. 
This is most easily done within the framework of the one boson exchange
approximation. The formalism and philosophy are discussed in \cite{jerry},
and the calculation is discussed in \cite{rmgm98}. The nucleon-nucleon
potential $V(NN)$ describes phase shifts reasonably well. The corresponding
density is
 ${\cal V}(NN)$. The basic Lagrangian density
 contains a free nucleon term ${\cal L}_0(N)$,
 a free meson term ${\cal L}_0({\rm mesons})$ and an interaction term
 ${\cal L}_I(N,{\rm mesons})$  but does not contain ${\cal V}(NN)$. Thus 
 one adds this term and subtracts it:
\bea{\cal L}&=&{\cal L}_0(N)-{\cal V}(NN) + {\cal L}_{\rm m}\\
 {\cal L}_{\rm m}& =&{\cal L}_I(N,{\rm mesons})+{\cal L}_0({\rm mesons})
+{\cal V}(NN).\eea
We  use
the term ${\cal L}_0(N)-{\cal V}(NN)$ to obtain a first solution
$\mid\Phi\rangle$ to the
many-body problem. The term 
${\cal L}_{\rm m} $
accounts for mesonic content of Fock space, and we
present\cite{rmgm98} a scheme to incorporate the 
effects of ${\cal L}_{\rm m} $ and calculate the full wave function
$\mid\Psi\rangle$. Our procedure allows us to assess whether or not
${\cal V}(NN)$ has been chosen well. If it has, the effects of 
${\cal L}_{\rm m} $ can be treated perturbatively.

Solving for $\mid\Phi\rangle$ is no easy task --it demands a separate
non-perturbative treatment. One introduces a mean field 
$U_{MF}$ which acts
on single nucleons.
\be{\cal L}_0(N)-{\cal V}(NN) ={\cal L}_0(N)-U_{MF} +
\left(U_{MF}-{\cal V}(NN)\right).\ee
The operator 
 $U_{MF}$ is chosen to minimize the effects of  $\langle \Psi|U_{MF}-{\cal
V}(NN)|\Psi \rangle.$
There is a well-known procedure, called Brueckner theory, which is used to
determine $U_{MF}$. In schematic terms: 
\be  U_{MF}\sim G  \times \rho,\ee
in which $G$ is a nucleon-nucleon scattering matrix, as modified by the Pauli
principle, $\rho$ is the nuclear density, and the $\times$ represents a
convolution.

The result \cite{rmgm98} is a rather complete theory in which the 
full wave function $|\Psi\rangle$ includes the
effects of both NN correlations and explicit
mesons.

\subsection*{Results }

The trivial nature of the vacuum in the light front formalism was exploited in
deriving\cite{rmgm98} the necessary equations.
       Applying our light front OBEP, the nuclear matter
       saturation properties are reasonably well reproduced. The
       binding energy per nucleon is 14.71 MeV with a value of $k_F$ of
       1.35 fm$^{-1}$. This is good considering that we have no three-body
       force. The computed
        value of the compressibility, 180 MeV,  is
       smaller than that of alternative relativistic approaches to nuclear
       matter in which the compressibility
       usually comes
       out too large.
      The  replacement of 
       meson degrees of freedom by a
       NN interaction was shown to be  a reasonable  approximation, and
       that the formalism allows one to
       calculate corrections to this approximation
       in a well-organized manner.
        The mesonic Fock space components of the
       nuclear wave function are studied we
       find that there are about 0.05 excess pions per nucleon.

The magnitudes of the scalar and vector potentials
are far smaller than found in the mean field approximation.
Our first calculation  
neglected the influence of
two-particle-two-hole states
to obtain an approximate version of 
$f(k^+)$ 
the  nucleons carry 81\% (as opposed to the 65\% 
of mean field theory) of  the nuclear plus momentum. 
This is a vast improvement in the description
of nuclear deep inelastic
scattering as the  
minimum value of the ratio $F_{2A}/F_{2N}$
 is increased by a factor of twenty   towards 
the data. This  is not enough to provide a satisfactory description, but it is
an excellent start.
I am optimistic about future
results because including nucleons
with momentum greater than $k_F$ can be expected to
substantially increase the
computed ratio $F_{2A}/F_{2N}$\cite{rmgm98}.

Let me discuss the observational aspects, concentrating on
the experimental information about the nuclear pionic content.
The Drell-Yan experiment on nuclear targets \cite{dyexp}
showed no enhancement of nuclear pions within an error of about 5\%-10\% for
their heaviest target. 
Understanding this result is 
an important challenge to the 
understanding of nuclear dynamics~\cite{missing}. 
Here we have a good description of nuclear dynamics, 
and our 5\%  enhancement is consistent,
within errors, with the Drell-Yan
data.

\section*{SUMMARY} 

The light front approach
has  been applied, within the mean field approximation,
to both infinite and finite  nuclear matter.
Furthermore, LF studies of 
$\pi N$ and $NN$ scattering have been made. This is
input to LF calculations of
correlated nucleons in
infinite nuclear matter.
One can use  light front dynamics  to compute nuclear
energies, wave functions and the experimentally observable  plus-momentum
distributions for a wide variety of Lagrangians.
There are indications that the computed quantities will
ultimately be in good agreement with experiment. 
The use of  light front dynamics in nuclear physics is only in its infancy,
but it seems to be a tool that can be used for any problem in high energy
nuclear physics.
\section*{Acknowledgments}
These lectures are based on work performed in collaboration with 
P.G.~Blunden, M.~Burkardt, and R.~Machleidt.

\end{document}